\documentstyle[preprint,aps]{revtex}

\begin{document}
\draft
\preprint{CMU-HEP93-19; DOE-ER/40682-44}
\title{Spontaneous CP Violation and Natural Flavor Conservation \\
In the $SU(2)_{L}\times U(1)_{Y}$ Gauge Theory with Two Higgs Doublets}
\author{Yue-Liang  Wu}
\address{ Department of Physics, \ Carnegie Mellon University \\ Pittsburgh,
 Pennsylvania 15213,\ U.S.A.}
\date{\today}
\maketitle

\begin{abstract}
 The two principles of spontaneous CP violation and natural flavor conservation
can be realized in the $SU(2)_{L}\times U(1)_{Y}$ model with two Higgs
doublets. In particular, this model provides
a consistent application to the CP violating parameters $\epsilon$ and
$\epsilon'/\epsilon$ in kaon decay and the neutron electric dipole moment.
The masses of the exotic scalars are unconstrained in this model and probing
these exotic scalars must be valuable at the present energy scales. Large CP
violations may also occur in the heavy quark and lepton sectors.
\end{abstract}

\narrowtext

   I find that the $SU(2)_{L}\times U(1)_{Y}$ gauge theory with two Higgs
doublets, which is the simplest extension of the standard
$SU(2)_{L}\times U(1)_{Y}$ model\cite{WSG},  can realize the two principles
that CP conservation is a fundemental symmetry of lagrangian and broken
spontaneously, and the neutral currents conserve all flavors naturally,
i.e., the usually so-called spontaneous CP violation (SCPV) which was first
pointed out by T.D. Lee\cite{TDLEE}, and the natural flavor conservation
(NFC) which was first suggested by Glashow and Weinberg\cite{GW}, and
Paschos\cite{EAP}. As a consequence, the indirect CP violation
$(\epsilon)$\cite{CCFT} and direct CP violation $(\epsilon'/\epsilon)$
\cite{FC} observed in kaon decay and the neutron electric dipole moment
(EDM)$(d_{n})$ \cite{SMITH} can be easily accounted for in this minimal
SCPV and NFC model with two Higgs doublets.

  The crucial point to achieve such a model is that the Glashow-Weinberg
criterion\cite{GW} for NFC in the Higgs sector is found to be only sufficient
but not necessary. I then deduce a new theorem of matrix algebra, instead of
the discrete symmetries imposed by the Glashow-Weinberg criterion,
to ensure NFC in the Higgs sector. Unlike the Glashow-Weinberg criterion, the
new theorem allows each Higgs doublet couple to all the fermions (leptons and
quarks) even if conforming to the NFC,  so that the lagrangian does
not possess any additional discrete symmetries. Glashow-Weinberg criterion
is only one of the special cases.

  Without assumming the NFC\cite{HALL}, the model becomes a
general SCPV two-Higgs-doublet model\cite{TDLEE}, and it usually has
flavor changing neutral Higgs exchanges (FCNH) which result in $\Delta S=2$
superweak interaction (SWI)\cite{WOLF} at tree level. Such a SCPV and
SWI model with accepting standard constraints on the Higgs mass was
first seriously investigated by Liu and Wolfenstein\cite{LIU}. They chosen
rather to have the unnatural heavy scalars\cite{BBG} in an $SU(2)\times U(1)$
model than to fine-tune the FCNH couplings to sufficiently small so that to
meet the restriction on $\Delta S=2$ given by the $K_{L}-K_{S}$ mass
difference. Consequently, they found that such a simple model provided
the abundant interesting phenomenological applications.

     Anyway we believe that the simplest extended $SU(2)_{L}\times U(1)_{Y}$
model with two Higgs doublets is an attractive and fascinating model.
I now present a detailed description for an alternative two-Higgs-doublet
model which has both the SCPV and the NFC.  Let us begin with the general
Yukawa interactions
\begin{eqnarray}
L_{Y} & = & \bar{q}_{L}\Gamma_{D}^{a}D_{R}\phi_{a} + \bar{q}_{L}\Gamma_{U}^{a}
U_{R}\bar{\phi}_{a} + \bar{l}_{L}\Gamma_{E}^{a}E_{R}\phi_{a} + h.c.
\end{eqnarray}
where $q_{i}$, $l_{i}$ and $\phi_{a}$ are $SU(2)_{L}$ doublet quarks,
leptons and Higgs bosons, while $U_{i}$, $D_{i}$ and $E_{i}$ are $SU(2)_{L}$
singlets, $\Gamma^{a}_{F}$ (F=U, D, E) are the Yukawa coupling matrices and
real by CP invariance. $i = 1,\cdots , n$ is a generation label and
$a = 1,2,\cdots , N$ is a Higgs doublet label.

      To ensure the NFC, the following theorem in the Higgs sector is
deduced to replace the Glashow-Weinberg criterion.

      Theorem: for the flavor conservation by the neutral currents to be
natural in the Higgs sector, or equivalently, the matrices
$\Gamma^{a}_{F}$ (F=U, D, E) $a=1,2, \cdots, N$ can be diagonalized
simultaneously by a biunitary or biorthogonal (for real $\Gamma^{a}_{F}$)
transformation if and only if the square $n\times n$ matrices
$\Gamma^{a}_{F}$ are represented in terms of
the linear combinations of a complete set of $n\times n$ matrices
$\{\Omega^{\alpha}_{F},\alpha =1,2, \cdots , n\}$, i.e
\begin{equation}
\Gamma_{F}^{a} = \sum_{\alpha} g_{a\alpha}^{F}\Omega_{F}^{\alpha}
\end{equation}
where $\Omega_{F}^{\alpha}$ satisfy the following orthogonal condition
\begin{equation}
\Omega_{F}^{\alpha}(\Omega_{F}^{\beta})^{\dagger} = L_{F}^{\alpha}
\delta_{\alpha\beta} \ ; \qquad (\Omega_{F}^{\alpha})^{\dagger}
\Omega_{F}^{\beta} = R_{F}^{\alpha}\delta_{\alpha\beta} \ .
\end{equation}
with the conventional normalization $\sum_{\alpha}L_{F}^{\alpha}
= \sum_{\alpha}R_{F}^{\alpha} = 1$.

    This theorem implies that the real matrices $\Gamma^{a}_{F}$ can be
written  into the following structure
\begin{equation}
\Gamma^{a}_{F} = \sum_{\alpha=1}^{n}g_{a\alpha}^{F} O_{L}^{F}\omega^{\alpha}
(O_{R}^{F})^{T} \ , \qquad a=1, \cdots ,N.
\end{equation}
with $\{\omega^{\alpha}, \alpha =1, \cdots,n\}$ the set of diagonalized
projection matrices $\omega^{\alpha}_{ij} = \delta_{i\alpha}
\delta_{j\alpha}$. $O_{L,R}^{F}$ are the arbitrary orthogonal matrices and
independent of the Higgs doublet label $a$, the later feature is the crucial
point for ensuring the NFC.  $g^{F}_{a\alpha}$ are the arbitrary real
Yukawa coupling constants.

   The proof of the above theorem is evident from the well-known theorem
of the matrix algebra that any matrix can be diagonalized by a biunitary
matrix. I prefer not to present a detailed demonstration on this theorem
in this short paper.

   Consider now the simplest case of two Higgs doublets, i.e., $N=2$.
After the spontaneous symmetry breaking, the neutral components of the Higgs
doublets acquire VEV's
\begin{equation}
<\phi_{1}^{0} > = v_1 e^{i\delta}/\sqrt{2} \ , \qquad <\phi_{2}^{0} > =
v_2 /\sqrt{2} \ .
\end{equation}
where a special phase basis is chosen so that $v_1$ and $v_2$ are real.
The mass matrices are given by the term
\begin{equation}
M_{F}=\sum_{a}\hat{v}_{a}\Gamma^{a}_{F} =\sum_{\alpha=1}^{n}(v_{1}e^{i\delta}
g_{1\alpha}^{F}+ v_{2}g_{2\alpha}^{F}) O_{L}^{F}\omega^{\alpha}(O_{R}^{F})^{T}
\end{equation}
Which are the arbitrary complex matrices. The physical basis is defined
through the orthogonal transformations and the phase redefinitions of the
fermions and scalars.
\begin{equation}
f_L = (O_{L}^{F})^{T}F_L \ ; \qquad  f_R = (O_{R}^{F}P^f)^{T}F_R \ .
\end{equation}
where $P^{f}$ is the diagonal matrix with the phase factor,
$P^{f}_{ij} = e^{i\sigma \delta_{f_{i}}}\delta_{ij}$, with $\sigma =+$,
for $f_i =d_i, e_i$, and  $\sigma =-$, for $f_i =u_i$.
The phases $\delta_{f_{i}}$ are introduced from the definition
\begin{equation}
(\sin\alpha g_{1i}^{F}e^{i\sigma\delta} +
\cos\alpha g_{2i}^{F})v \equiv \sqrt{2}m_{f_{i}}e^{i\sigma\delta_{f_{i}}}
\end{equation}
with $v^2 = v_{1}^{2} + v_{2}^{2}= (\sqrt{2}G_{F})^{-1}$, $\sin\alpha
= v_1/v$ and $\cos\alpha = v_2/v$. Where $m_{f_{i}}$ are the masses of the
physical states $f_{i}$. We may call $\delta_{f_{i}}$ the induced phases from
the complex VEV's.

    In the physical basis, the Yukawa interaction term becomes
\begin{eqnarray}
L_{Y} & = &
(2\sqrt{2}G_{F})^{1/2}\sum_{i,j=1}^{3}[\xi_{d_{j}}\bar{u}_{L}^{i}V_{ij}
m_{d_{j}}d^{j}_{R}H^+ - \xi_{u_{j}}^{\ast}\bar{d}_{L}^{i}V_{ij}^{T}
m_{u_{j}}u^{j}_{R}H^-  \nonumber \\
 & & + \xi_{e_{j}}\bar{\nu}_{L}^{i}\delta_{ij}
m_{e_{j}}e^{j}_{R}H^+ + h.c. ] + (\sqrt{2}G_{F})^{1/2}\sum_{i=1}^{3}
\sum_{k=1}^{3} [\eta_{u_{i}}^{(k)\ast}m_{u_{i}}\bar{u}_{L}^{i}u_{R}^{i} \\
 & & + \eta_{d_{i}}^{(k)}m_{d_{i}}\bar{d}_{L}^{i}d_{R}^{i} +
 \eta_{e_{i}}^{(k)}m_{e_{i}}\bar{e}_{L}^{i}e_{R}^{i} + h.c.]H_{k}^{0} \nonumber
\end{eqnarray}
with
\begin{equation}
 \xi_{f_{i}} = \frac{2\sin\delta_{f_{i}}}
{\sin2\alpha \sin\delta}e^{i(\delta - \delta_{f_{i}})} - \tan\alpha \  ;
\qquad  \eta_{f_{i}}^{(k)} = O_{2k}e^{-i\delta_{f_{i}}} + (O_{1k}+iO_{3k})
\xi_{f_{i}} \ .
\end{equation}
where $V=(O^{U}_{L})^{T}O^{D}_{L}$ is the real Cabbibo-Kobayashi-Maskawa
matrix. $H^{0}_{k}=(h, H, A)$ are the three physical neutral scalars
and $O_{kl}$ is the $3\times 3$ orthogonal mixing matrix among these three
scalars, $H$ plays the role of the Higgs boson in the standard model.

The induced complex fermion-Higgs boson Yukawa couplings are given by
\begin{equation}
 g_{f_{i}} \equiv (2\sqrt{2}G_{F})^{1/2}\xi_{f_{i}}m_{f_{i}} =
 g_{1i}^{F}\arccos\alpha \ e^{i(\delta - \delta_{f_{i}})} -
(\sqrt{2}G_{F})^{1/2} m_{f_{i}} \tan\alpha
\end{equation}

  Without making additional assumptions, $m_{f_{i}}$, $V_{ij}$,
$m_{H_{k}^{0}}$, $\xi_{f_{i}}$(or $g_{f_{i}}$, or $\delta_{f_{i}})$,
$m_{H^{+}}$ and $O_{kl}$ (or $\eta_{f_{i}}^{(k)}$) are in general the free
parameters. $m_{f_{i}}$, $V_{ij}$, and $m_{H^{0}}$ already appear in
the standard model with one Higgs doublet. The additional parameters
$\xi_{f_{i}}$ and $O_{kl}$ (or $\eta_{f_{i}}^{(k)}$) should measure the
magnitudes which deviate from ones in the standard model, and describe the
new physical phenomena.

     An important feature of the present model is that the structure of the
fermion-Higgs boson Yukawa couplings in eq.(9) distinguishes from one in the
Weinberg three-Higgs-doublet model\cite{SW1}  in which the complex Yukawa
couplings (corresponding to the $\xi_{f_{i}}$) are the same for the
fermions with given charge and depend only on the Higgs bosons. This is
because the CP violation of the charged-Higgs boson sector in the Weinberg
model originates from the complex mass matrix of the charged-Higgs bosons,
for which three Higgs doublets are the minimal number required for a nontrivial
CP phase in the charged-Higgs boson mixing matrix.  Therefore the CP
violations occured in the processes which involve the  different flavors of
the same charge are correlated strongly each other in the Weinberg model of
CP violation. It is this essential reason that why the Weinberg
three-Higgs-doublet model suffers from the inconsistencies between the
constraints obtained from K-physics and from the neutron EDM as well as
B-physics\cite{SW2,HYC}. Unlike the Weinberg model of CP violation, the
induced CP violating phases $\delta_{f_{i}}$ (or the corresponding complex
Yukawa couplings $\xi_{f_{i}}$) of the present model arise from the
redefinition of the phases of the quarks and leptons after the spontaneous
symmetry breaking, and they are all different and can not be rotated away,
so that they become nontrivial and observable in the fermion-Higgs boson
interactions. The basic reason for such a case is because all the quarks and
leptons in the present model can couple to the two Higgs bosons, and receive
their contributions to their mass from the two VEV's and the corresponding two
Yukawa couplings which are different for all the fermions. Obviously, these
induced phases also contribute to the fermion-neutral Higgs boson couplings
$\eta_{f_{i}}^{(k)}$. Moreover, there are additional CP violating sources
which contribute to the  $\eta_{f_{i}}^{(k)}$. The origin of these sources is
well-known due to the scalar-peseudoscalar mixing. As a consequence, the
present model evades the difficulties encounted in the Weinberg
three-Higgs-doublet model. To see this, let us examine this simple model in
its phenomenological applications.

   Consider first the CP violating parameters $\epsilon$ and
$\epsilon'/\epsilon$ in kaon decays. Like the Weinberg
three-Higgs-doublet model, but unlike the Kobayashi-Maskawa model\cite{KM},
the long-distance contribution to $\epsilon$ dominates
over the short-distance one. The imaginary part of the $K^0-
\bar{K}^0$ mixing mass matrix to which $\epsilon$ is propotional mainly
receives the contribution from the $\pi$, $\eta$ and $\eta'$ poles\cite{HAG}.
Using the estimates given in the literature\cite{HYC} for the hadronic
matrix elements, I find that $\epsilon$ can be easily accounted for
by taking
\begin{equation}
 Im(\xi_s\xi_{c}^{\ast}-0.05\xi_{d}^{\ast}\xi_c)\frac{1}{m_{H^{+}}^{2}}
(ln\frac{m_{H^{+}}^{2}}{m_{c}^{2}}-\frac{3}{2})
\simeq 2.7\times 10^{-2} GeV^{-2}
\end{equation}
where the experimental value $|\epsilon| = 2.27\times 10^{-3}$ has been used.

   For the ratio $\epsilon'/\epsilon$, the calculations evaluated in
Ref.\cite{HYC,DH} can be applied to the prsent  model. In fact, the ratio does
not
depend on the detail of the CP-odd matrix element, so that their evaluations
for the ratio  $\epsilon'/\epsilon$ are also valid for the
present model.  Within the theoretical uncertainties, the ratio is reanalysed
recently to be $\epsilon'/\epsilon = (0.4-6.0)\times 10^{-3}$\cite{HYC},
which is comparable to one calculated from the KM model\cite{YLWU}
and is also consistent with the present experimental data\cite{FC}.

     Consider now the neutron EDM, $d_{n}$. The present experimental limit
on $d_{n}$ is\cite{SMITH} $d_{n} < 1.2 \times 10^{-25} e cm$. Applying various
well-known scenarios for the calculations of $d_{n}$ to the present
model, the following constraints on parameters $\xi_{f_{i}}$ and
$\eta_{f_{i}}^{(k)}$ are obtained.

      In the quark model through charged-Higgs boson exchange
 \begin{equation}
 Im(\xi_d\xi_{c}^{\ast})\frac{1}{m_{H^{+}}^{2}}
(ln\frac{m_{H^{+}}^{2}}{m_{c}^{2}}-\frac{3}{4})
\alt 6.3\times 10^{-2} GeV^{-2}
\end{equation}

     The Weinberg's  gluonic operator through charged-Higgs boson
exchange\cite{DICUS} leads to
 \begin{equation}
 Im(\xi_b\xi_{t}^{\ast})h_{CH}(m_{t},m_{b},m_{H})
\alt 3.0\times 10^{-2}
\end{equation}
where $h_{CH}$ is a function of the quark- and Higgs-mass arising from the
integral of the loop. $h_{CH}\leq 1/8$ and $h_{CH}=1/12$ for $m_{b}\ll
m_{t}=m_{H}$. For the later case, one has  $Im(\xi_b\xi_{t}^{\ast})\leq 0.36$.

 From the neutral-Higgs boson exchange\cite{SW2}, it requires
 \begin{equation}
 Im [\eta_{t}^{(k)\ast}]^{2} h_{NH}(m_{t},m_{H_{k}^{0}}) \alt 0.18
\end{equation}
with $h_{NH}$ a similar function as $h_{CH}$. $h_{NH}\simeq 0.05$ for
$m_{t}=m_{H}$.

  The most dominant contribution to $d_{n}$ was found from the gluonic
chromoelectric dipole moment (CEDM)\cite{DWC} induced by the Barr-Zee two-loop
mechanism\cite{BZ}. For $m_{t}\sim m_{H}$, it puts a constraint
 \begin{equation}
 Im(\eta_{d}^{(k)} \eta_{t}^{(k)\ast} + 0.5\eta_{u}^{(k)\ast}
\eta_{t}^{(k)\ast}) \alt 0.2
\end{equation}

  Combining  the result obtained from fitting the $\epsilon$ and the constraint
from accomodating the $d_{n}$ in the quark model,  it is not difficult to
find, for the present experimental lower limit of $m_{H^{+}}\agt 45$GeV, that
\begin{equation}
Im\xi_{s}\xi_{c}^{\ast} \agt 10 \ ; \qquad Im\xi_{d}\xi_{c}^{\ast}\sim 21
\end{equation}
which shows that at least one of the Yukawa couplings should be larger
than the one in the standard model. A natural solution is
$\sin\alpha \ll 1$, i.e., $v_{1} \ll v_{2}$, which implies that the
complex fermion-Higgs boson Yukawa couplings $g_{f_{i}}$
$(f_{i}=u_{i},d_{i},e_{i})$ can, in general,  be larger than the ones
in the standard model. The above constraints do not exclude
the possible large CP violations in the heavy quark system due to
the large complex Yukawa couplings.

    Finally, I would like to make the following remarks. First,
the domain-wall problem is absent from any additional discrete symmetries
except the CP symmetry which is encounted by most SCPV models, it is also
supposed to ignore this problem in the present paper. Second, the strong CP
problem may be simply evaded by using the well-known Peccei-Quinn mechanism
\cite{PQ} realized in a heavy fermion invisible axion scheme\cite{KSVZ}.
Since this scheme, on the one hand, is one of the simplest schemes and has an
advantage that it leaves the above model unchanged except to regard the
orthogonal matrix $O_{kl}$ as a $4\times 4$ matrix to include the small
mixings among the additional singlet scalar and the three physical neutral
Higgs bosons, and on the other hand it is also free from the axion domain-wall
problems.  The third, although CP violation solely originates from the
spontaneous symmetry breaking, the mechanism to have CP violation to take
place in the lagrangian after the spontaneous symmetry breaking can be
different in various models. The examples are the T.D. Lee  model\cite{TDLEE},
the Weinberg model\cite{SW1}, the Liu-Wolfenstein model\cite{LIU} and the
model described in this paper. The fourth, a consistent phenomenological
application of the present model may imply that the FCNH in a general
two-Higgs-doublet model can be made to be very small and even negligible,
so that the SWI through the neutral-Higgs boson exchange still keeps its
generic features described from the original motivation by Wolfenstein
\cite{WOLF}, otherwise it may deviate significantly from a generic superweak
model due to an unexpected large value of $\epsilon'/\epsilon$ \cite{LIU}.
Last but not least,  the present model leaves  the masses of the exotic
scalars unconstrained, which strongly indicates that probing these exotic
scalars must be valuable at the present energy scale.  Large CP violations
in the heavy flavor and lepton sectors may also open a window to the
experimental detections.

\acknowledgments

   I am grateful Lincoln Wolfenstein for valuable discussions which have
motivated my intention from three Higgs doublets to the simplest two Higgs
doublets model and for useful suggestions in the preparation of this paper.
I also wish to thank K.C. Chou for having inspired my interest in the SCPV,
and E.A. Paschos for valuable conversations on the NFC,  Richard P. Holman
and Ling-Fong Li for useful discussions on the domain-wall problem. I have
also benefitted from conversations with Luis Lavora,  Martin Savage and
Joao P. Silva. This work was supported by DOE grant \# DE-FG02-91ER40682.


\begin{thebibliography}{99}
\bibitem{WSG} S. Weinberg, Phys. Rev. Lett. 19, 1264 (1967); A. Salam, in
{\it Elementary Particle Physics}, edited by N. Svartholm (Almquist and
Wikselle, Stockholm, 1968).
\bibitem{TDLEE} T.D. Lee, Phys. Rev. D8, 1226 (1973); Phys. Rep. 9, 143 (1974).
\bibitem{GW} S.L. Glashow and S. Weniberg, Phys. Rev. D15, 1958 (1977);
\bibitem{EAP} E.A. Paschos, Phys. Rev. D15, 1966 (1977).
\bibitem{CCFT} J.H. Christenson, J.W. Cronin, V.L. Fitch and R. Turlay,
Phys. Rev. Lett. 13, 138 (1964).
\bibitem{FC} M. Burkhardt, et al., Phys. Lett. B237, 303 (1990);
J.R. Patterson, Phys. Rev. Lett. 64, 1491 (1990);
For a recent review on theoretical and experimental progresses, see,
for example, B. Winstein and L. Wolfenstein, CMU-D.O.E. ER/40682-24,
EFI 92-55, to be published in Rev. Mod. Phys. 1993.
\bibitem{SMITH} K.F. Smith et al., Phys. Lett. B234, 234 (1990).
\bibitem{HALL} Recently, it was shown that by considering approximate flavor
symmetries, the exotic Higgs bosons may be allowed to have a weak-scale
mass (few hundrets GeV) so that the Glashow-Weinberg criterion is rendered
unnecessary, nevertherless, one has to be very careful to choose the
Yukawa couplings. A. Antaramian, L.J. Hall and A. Rasin, Phys. Rev. Lett. 69,
1871 (1992); L.J. Hall and S. Weinberg, Phys. Rev. D48, 979 (1993); A. Rasin
and J.P. Silva, LBL-34592; UCB-PTH-93/24, CMU-HEP93-11, 1993.
\bibitem{WOLF} CP violation through superweak interaction was first proposed
by L. Wolfenstein, Phys. Rev. Lett. 13, 562 (1964). An important prediction in
this theory is $\epsilon'/\epsilon\simeq 0$.
\bibitem{LIU} J. Liu and L. Wolfenstein, Nucl. Phys. B289, 1 (1987).
\bibitem{BBG} G.C. Branco, A.J. Buras and J.M. Gerard, Nucl. Phys. B259,
306 (1985).
\bibitem{SW1} S. Weinberg, Phys. Rev. Lett. 37, 657 (1976).
\bibitem{SW2} S. Weinberg, Phys. Rev. Lett. 63, 2333 (1989), and references
therein; Phys. Rev. D42, 860 (1990).
\bibitem{HYC} H.Y. Cheng, Int. J. Mod. Phys. A7, 1059 (1992), and references
therein.
\bibitem{KM} M. Kobayashi and T. Maskawa, Prog. Theor. Phys. 49, 652 (1973).
\bibitem{HAG}  Y. Dupont and T.N. Pham, Phys. Rev. D28, 2169 (1983); D. Chang,
Phys. Rev. D25 (1982) 1318; J. Haglin, Phys. Lett. B117, 441 (1982).
\bibitem{DH} J.F. Donoghue and B.R. Holstein, Phys. Rev. D32, 1152 (1985);
H.Y. Cheng, Phys. Rev. D34, 1397 (1986).
\bibitem{YLWU} See, for example, Y.L. Wu,  {\it Recent theoretical development
on direct CP violation $\epsilon'/\epsilon$}, in: Proc. of the XXVI Int..
Conf. on High Energy Phyics, p. 506, 1992-Dallas, Texas, edited by
J.R. Stanford, and references therein.
\bibitem{DICUS} D.A. Dicus, Phys. Rev. Lett. 41 999 (1990).
\bibitem{DWC} D.W. Chang, W.Y. Keung and T.C. Yuan, Phys. Lett. B251,
608 (1990); J. Gunion and D. Wyler, Phys. Lett. B248, 170 (1990).
\bibitem{BZ} S.M. Barr and A. Zee, Phys. Rev. Lett. 65, 21 (1990).
\bibitem{PQ} R.D. Peccei and H.R. Quinn, Phys. Rev. Lett. 38, 1440 (1977);
Phys. Rev. D16, 1791 (1977).
\bibitem{KSVZ} J.E. Kim, Phys. Rev. Lett.43, 103 (1979); M.A. Shifman,
V.I. Vainstein and V.I. Zakharov, Nucl. Phys. B166, 4933 (1980).
\end{thebibliography}
\end{document}